\documentclass{article}		
\usepackage{latexsym}		  

\def\AUTHOR{William M. Pezzaglia Jr.}
\def\EMAIL{wpezzag@clifford.org}
\def\HTTP{ Http://www.clifford.org/wpezzag/talk/99mexico/ }
\def\TITLE{Polydimensional Supersymmetric Principles}

\begin{document}
\font\sixsy=cmsy6

\def\NOTES#1{{\tt #1}}                
\def\HBAR{{\mathchar'26\mkern-9muh}}  
\def\vector#1{\vec{\bf {#1}}}         
\def\norm#1{\parallel{#1}\parallel}   
\def\LLL{{\cal L}}                    
\def\EQN#1{eq.\ (#1)}                 
\def\EB#1{{\bf e}_{#1}}
\def\EBH#1{{\bf \hat{e}}_{#1}}
\def\HALF{\mathstrut{1 \over 2}}                
\def\ODOT#1{\stackrel{\circ}{{#1}}}   
\def\ODOTS#1#2{\ODOT{#1}{\!}^{#2}}
\def\ODOTL#1#2{\ODOT{#1}{\!}_{#2}}
\def\VARO#1#2{\delta\!\!\ODOT{#1}{\!}^{#2}}

\def\COV#1{\nabla\!_{#1}}             
\def\PV#1#2{{\delta {#1} \over \delta {#2} }}
\def\PD#1#2{{\partial {#1} \over \partial {#2}}}
\def\PP#1#2#3{{\partial^2 {#1} \over \partial {#2} \partial {#3}}}
\def\BD#1#2{{\left[\partial_{#1},\partial_{#2}\right]}}
\def\CD#1#2{\left[ \partial_{#1}, \partial_{#2} \right]}
\def\LLL{{\cal L}}

\begin{titlepage}

\title{ \Large\bf \TITLE
\thanks{ Based on presentation available at \HTTP } }
\author{ {\Large\bf  \AUTHOR }
\thanks{Email:\EMAIL} \\
\\ Department of Physics
\\ Santa Clara University
\\ Santa Clara, CA 95053 }
\maketitle
\thispagestyle{empty}

\begin{abstract}
Systems of equations are invariant under {\it polydimensional
transformations}\/ which  reshuffle the geometry such that what
is a line or a plane is dependent upon the frame of reference.
This leads us to propose an extension of Clifford calculus in which
each geometric element (vector, bivector) has its own coordinate.
A new classical action principle is proposed in which particles take
paths which minimize the distance traveled plus area swept out by the
spin.  This leads to a solution of the 50 year old conundrum of `what
is the correct Lagrangian' in which to derive the Papapetrou equations
of motion for spinning particles in curved space (including torsion).

{\it Based on talk given at \/}{\bf 5th International Conference
on Clifford Algebras and their Applications in Mathematical Physics,}
{\it Ixtapa-Zihuatanejo, Mexico, June 27-July 4, 1999.\/}
{\it Submitted to International Journal of Theoretical Physics.}
\end{abstract}

\end{titlepage}

\pagestyle{myheadings}

\section*{1. Introduction}
Reformulating physical laws with a new mathematical language will not
in itself lead to new principles.  However, because Clifford algebra
(Lounesto, 1997)
encodes the structure of the underlying geometric space, we see
possible bigger patterns emerge.  Specifically in the description
of a spinning particle the equations of motion are invariant
under a non-dimensional preserving {\it polydimensional}\/
transformation which rotate between vector momentum and
bivector spin.
This leads us to propose that the physical laws might be covariant
under general automorphism transformations which reshuffle the
geometry, a classical analogy to the quantum model of
Crawford (1994).

The invariants of these transformations suggest
that the spin motion contributes to the proper time.
Hence a new action principle is proposed in which particles take
paths which minimize the sum of the linear distance traveled
combined with the bivector area swept out by the spin.
In curved space, the velocity of the variation is not the variation
of the velocity, leading to a new derivation of the Papapetrou (1951)
equations (of a spinning particle) as autoparallels in the 
polydimensional space.

\section*{2. Relative Dimensionalism}
Is the dimension of the geometric quantity (e.g. scalar, vector)
absolutely unique to the associated physical quantity?  Certainly
mass is a 0D (zero-dimensional) scalar while momentum a 1D vector.
In contrast, consider that while time is viewed as the 4th
dimension in Minkowski space, special relativity was originally
formulated with time treated as a scalar.  Is there a right/wrong answer
as to the geometric nature of time, or is it a function of the
observer's frame of reference?  We suggest that {\it dimension is
relative\/}(Pezzaglia, 1998a),
such that we can consider transformations which reshuffle
the basis geometry (e.g. vector line replaced  by bivector plane),
yet leave sets of physical laws invariant.  One
application provides a new derivation of the enhancement of the 
mechanical mass by the amount of spin.

\subsection*{2.1.  Review of Special Relativity}
In electrodynamics one can unify a 3D vector force law with a
3D scalar work-energy law,
$$\dot{{\cal E}} = e \vector{E} \bullet \vector{v}\ , \eqno(1)$$
$$\dot{\vector{P}} = e \left( \vector{E} + \vector{v}\times
\vector{B} \right)\ , \eqno(2)$$
into one single equation,
$$\dot{p}^\mu = \left({e \over m}\right) p_\nu F^{\mu\nu}\ ,
\eqno(3)$$
using 4D vectors (and tensors).
Certainly the adoption of the four-dimensional
viewpoint has notational economy, and provides insight that the
work-energy theorem (1) is simpy the fourth aspect of the vector
force law (3).  However, philosophically one can ask if the 4D viewpoint
is any more correct than the 3D equations as they describe the
same phenomena.  Since special relativity
was originally formulated without the concept of Minkowski spacetime,
it is convenient, but apparently not necessary to
adopt the paradigm shift from 3D to 4D.
Hence we are being purposely dialectic in raising the question 
whether one can make an absolute statement about the dimensional
nature of a physical quantity such as time.  Can we state
(measure) that time is a part of a four-vector (as opposed to
a 3D scalar), or is this relative to whether one adopts a 3D or
4D world view, hence relative to the observer's dimensional 
frame of reference?

Usually the behavior of quantities under symmetry transformations
is used to define the dimensional nature (e.g.  does a set of
three quantities transform like a vector under rotations).   
In classical physics the fundamental laws must be invariant under
rotational displacements because it is postulated that the universe
is {\it isotropic} (has no preferred direction).
When one formulates laws with vectors (which are inherently coordinate
system independent),
isotropy is `built in' without needing to separately impose the
condition.  Hence (Gibbs) vectors are a natural language to express
classical (3D) physical laws because they naturally encode isotropy.
Einstein further postulated the metaprinciple that motion was relative;
that there is no absolute preferred rest frame to the universe.  This
coupled with the postulate that the speed of light is the same for all
observers leads to the principle that the laws of physics must be invariant
under Lorentz transformations (which connect inertial frames of reference).
As a consequence, in a 3D perspective, what is pure scalar (e.g. time
interval) to one observer is part scalar, part vector to another
observer.  Lorentz transformations, which are rotations in 4D spacetime
that preserve the dimension of the geometry,
in a 3D viewpoint NOT dimensional preserving.

In 3D space the length (magnitude) of a vector (e.g. electric field
or momentum) is invariant under rotations.  Under Lorentz transformations
(4D rotations), the modulus of the four-vector is invariant,
$$\norm{{\bf p}}^2 \equiv p_\mu p^\mu = 
{\cal E}^2/c^2 -\norm{\vector{P}}^2\ . \eqno(4)$$
Reinterpreted with a 3D viewpoint, the invariant quantity is the
difference between the square of the scalar energy minus the magnitude
of the 3D momentum vector.
Neither the modulus of the 3D scalar energy, nor 3D vector momentum is
independently invariant under these transformations.  Further, in
the 3D viewpoint it is as if the mass of the particle
(e.g. in definition of momentum: $p=mv$) has been increased by
its kinetic energy content,
$$m \equiv m_0 
\sqrt{1 + \left({\norm{\vector{P}} \over m_0 c}\right)^2}\ . \eqno(5)$$

\subsection*{2.2.  Automorphism Invariance}
Physicists usually first encounter Clifford algebras in quantum mechanics
in the form of Pauli, Majorana and Dirac `spin' matrices.  The spin-space
analogy to isotropy is that the physical formulation must be covariant
under global rotations of the spin basis.  An equivalent metaprinciple
would be to require that the physics is invariant under a change of
representation of the Dirac matrices.

Its possible however to avoid talking about the matrix representation 
entirely.  The more general concept is an algebra automorphism, which
is a transformation of the basis generators $\gamma_\mu$ of the
algebra which preserves the Clifford structure,
$$\{ \gamma_\mu , \gamma_\nu \} = 2 g_{\mu\nu}\ , \eqno(6) $$
where $g_{\mu\nu}$ is the spacetime metric.  For example, 
consider the following orthogonal transformation on any
element $\Gamma$ of the Clifford algebra,
$$\Gamma^{\prime} = {\cal R}\>\Gamma\>{\cal R}^{-1}\ , \eqno(7)$$
$${\cal R}(\phi)\equiv \exp(\gamma_\mu\>\phi^\mu /2)\ ,
\ \ \mu=1,2,3,4.  \eqno(8)$$
Proposing local covariance of the Dirac equation under such
automorphism transformations is one path to gauge theories of gravity
(see Crawford, 1994).

If the elements $\gamma_\mu$ are interpreted geometrically as basis
vectors, then (8) reshuffles geometry.  For example, when $\phi^4=
\pi /2$, equation (7) causes the permutation,
$$\gamma_j \Longleftrightarrow \gamma_4 \gamma_j\ ,
\ \ \ j = 1,2,3, \eqno(9)$$
$$\gamma_1 \gamma_2 \gamma_3 \Longleftrightarrow
\gamma_4 \gamma_1 \gamma_2 \gamma_3 \ , \eqno(10)$$
which exchanges three of the vectors with their associated timelike
bivectors.  What is a 1D vector in one ``reference frame'' is hence a
2D plane in another.  The transformation (8) thus ``rotates''
vectors into planes.

\subsection*{2.3.  Polydimensional Formulation}
Just as four-vectors allowed us to unify two equations into
one, the language of Clifford algebra allows for further notational
economy.  Consider that a classically spinning charged particle
obeys the torque equation of motion,
$$\dot{S}^{\mu\beta} = \left( {e \over m} \right) \left(
F^\mu_{\ \ \nu}\> S^{\nu\beta} - F^\beta_{\ \ \nu} \>
S^{\nu\mu} \right)\ . \eqno(11)$$
This and (3) can be written in the single statement,
$$\dot{\cal M} = \left( { e \over 2m } \right) \left[
{\cal M}, {\bf F} \right]\ , \eqno(12) $$
where ${\bf F} = {1 \over 2} F^{\mu\nu}\, \EB{\mu} \wedge \EB{\nu} $
is the electromagnetic field bivector and $\EB{\mu}$ are the basis
vectors of the geometric space.  The {\it momentum
polyvector\/} (Pezzaglia, 1998b) 
is defined as the multivector sum of the
vector linear momentum and the bivector spin momentum,
$${\cal M} \equiv p^\mu\>\EB{\mu} + {1 \over 2 \lambda}
\>S^{\mu\nu}\>\EB{\mu} \wedge \EB{\nu}\ , \eqno(13)$$
where $\lambda$ is some fundamental length scale constant (to be
interpreted in the next section).
The ability to add different ranked (dimensional) geometries
is the notational advantage of Clifford geometric algebra over
standard tensors.
Mathematically, (12) allows one to simultaneously obtain solutions
to both equations (3) and (11).

It is interesting to note that (12) is invariant under the
automorphism transformations generated by (8).  For example,
example, $\phi^4=\pi /2$ in (9) causes a trading between momentum and
mass moment of the spin tensor,
$$\lambda \> p_j \Longleftrightarrow  S_{4j}\ . \eqno(14)$$
It is not at all clear what physical interpretation to ascribe to the
two frames of reference.  A radical assertion of the {\it principle
of relative dimensionalism\/} (Pezzaglia, 1998a) would be to propose that
what is a vector to one observer is a bivector to another, and that 
they would partition the polymomentum (13) into momentum and spin
portions differently.  What is spin to one would be momentum to
the other.

Under the general rotation of the vectors into bivectors, both
observers would agree that the following generalized modulus
of the polyvector (13) would be invariant,
$$\norm{ {\cal M}}^2 \equiv \> p_\mu\>p^\mu
+\lambda^{-2}\>S_{\mu\nu}\>S^{\nu\mu}\ . \eqno(15)$$
In the $(---+)$ metric signature we define the modulus to be
the {\it bare mass}\/: $m_0 \equiv c^{-1} \norm{\cal M}$.
This implies that the mechanical mass (modulus of the linear momentum)
is NOT invariant under these transformations, but has been
enhanced by the spin energy content,
$$m \equiv c^{-1} \norm{{\bf p}} = m_0\  \sqrt{1 + 
{S^{\mu\nu}\>S_{\mu\nu} \over 
\left( m_0\>c\>\lambda \right)^2}}\ , \eqno(16)$$
in analogy to (5).  What we have described in (15), by simple geometric
construction, is a familiar result, laboriously obtained by Dixon
(1970) in the mechanical analysis of spinning bodies.
Expanding (16) non-relativistically one sees that $\lambda$
is consistent with the {\it radius of gyration} of a classical
extended particle.

\section*{3.  New Action Principle}
The polymomenta gives the (vector) linear momenta and (bivector)
spin momenta equal importance.  We now propose that each
quantity democratically has its own conjugate coordinate.
The generalized action principle is that particles take the paths
which minimize the sum of the linear distance traveled combined with
the bivector area swept out.  This simple geometric idea gives a new
derivation of the spin enhanced mass described by the Dixon
equation (15) and the Weysenhoff condition for spinning particles.

\subsection*{3.1.  Review of Classical Mechanics}
Classical
particles will follow paths of least spacetime distance between
endpoints, even when the space is curved by gravity.  The measure
of distance between two points in flat spacetime is,
$$c^2 d\tau^2 \equiv c^2 dt^2 - \left(dx^2 + dy^2 + dz^2
\right) = dx^\alpha\>dx^\beta\>g_{\alpha\beta}\ , \eqno(17)$$
where affine parameter $\tau$ is commonly called the {\it proper time}.
If we adopt the 3D viewpoint, we are combining (in quadrature)
the `scalar' time displacement with the `vector' path displacement,
utilizing a fundamental constant $c$ (the speed of light)
to combine the unlike quantities.

To obtain the equations of motion, one minimizes (extreemizes) the
{\it action integral}, which is based upon the quadratic form (17),
[note $x^4 \equiv ct$],
$${\cal A} \equiv \int {\cal L}\ d\tau = 
\int m_0 c\ d\tau = 
\int m_0 c\ \sqrt{\dot{x}^\alpha\>\dot{x}^\beta
\>g_{\alpha\beta} } \ d\tau\ . \eqno(18)$$
The integrand ${\cal L}$ is called the {\it Lagrangian}, which is 
generally a function of the coordinates $x^\alpha$,
and the velocities: $\dot{x}^\alpha = dx^\alpha / d\tau$
relative to the proper time.

The canonical  {\it four-momentum} $p_\mu$ is defined,
$$p_\mu \equiv {\delta {\cal L\ } \over \delta u^\mu} = m_0 u_\mu
=m_0 \dot{x}_\mu\ , \eqno(19) $$
which obeys (4).  It is easy to show that the 3D part of the
momentum $P_j=m v_j$ has mass $m$ which is enhanced by the
energy content according to (5).

\subsection*{3.2.  Dimensional Democracy}
If we fully embrace the concept of {\it relative dimensionalism},
then we must recognize that what one observer labels as a `point'
in spacetime with vector coordinates $(t,x,y,z)$ may be seen as an
entirely different geometric object by another.  This suggests that
perhaps we should formulate physics in a way which
is completely {\it dimensionally democratic\/} (Pezzaglia, 1998b) in
that all ranks of geometry are equally represented.  We propose 
therefore that `the world' is not the usual four-dimensional manifold,
but instead a fully {\it polydimensional continuum}, made of points,
lines, planes, etc.  Each event $\Sigma$ is a geometric point in a
{\it Clifford manifold}\ (Chisholm and Farwell, 1991),
which has a coordinate $q^A$ associated with each
basis element ${\bf E}_A$ (vector, bivector, trivector, etc.).  The 
{\it pandimensional differential} in the manifold would be,
$$d\Sigma \equiv {\bf E}_A dq^A=\EB{\mu}dx^\mu +
{1 \over 2 \lambda} \EB{\alpha}\wedge\EB{\beta}\>da^{\alpha\beta}
+{1 \over 6 \lambda^2} \EB{\alpha}\wedge\EB{\beta}\wedge\EB{\sigma}
\>dV^{\alpha\beta\sigma} +\dots\>, \eqno(20)$$
where in Clifford algebra it is perfectly valid to add vectors 
to planes and volumes (parameterized by the antisymmetric tensor
coordinates $dx^\mu,
da^{\alpha\beta}, dV^{\alpha\beta\sigma}$ respectively).

In analogy to (15), we propose that the quadratic form of the 
Clifford manifold would be the scalar part of the square of (20)
$$\norm{d\Sigma}^2 \equiv dx^\mu dx_\mu +
{1 \over 2 \lambda^2} da^{\alpha\beta} da_{\beta\alpha}
+ {1 \over 6 \lambda^4}dV^{\alpha\beta\sigma}
dV_{\sigma\beta\alpha} +\dots\ . \eqno(21)$$
The fundamental length constant $\lambda$ must be introduced in order
to add the bivector `area' coordinate contribution to the vector
`linear' one (Pezzaglia, 1998b).  This suggests that we have a new 
affine parameter $d\kappa=\norm{d\Sigma}$ which we will use to
parameterize our `polydimensional' equations of motion.

Classical mechanics assumes point particles that trace out linear paths.
The equations of motion are based upon minimizing the distance of
the path.  String theory introduces one-dimensional objects which
trace out areas, and the equations of motion are analogously
based upon minimizing the total area.  Membrane theory proposes
two-dimensional objects which trace out (three-dimensional)
volumes to be minimized.  Our new action principle suggests
that we should add all of these
contributions together, and treat particles as polygeometric objects
which trace out polydimensional paths with (21) the quantity to be
minimized.

\subsection*{3.3.  Application to the Classical Spinning Particle}
Using only the vector and bivector contributions of (21) the
Lagrangian that is analogous to (18) would be,
$${\cal L}(x^\alpha,\ODOTS{x}{\alpha},a^{\alpha\beta},
\ODOTS{a}{\alpha\beta} )= m_0 c
\>\sqrt{\ODOTS{x}{\mu} \ODOTS{x}{\nu}\>g_{\mu\nu}+{1 \over 2\lambda^2}
\ODOTS{a}{\alpha\beta}\ODOTS{a}{\mu\nu}\>g_{\beta\mu}\>g_{\alpha\nu}
}\ , \eqno(22)$$
where the open dot denotes differentiation with respect to the new
affine parameter (whereas the small dot is with respect to the proper
time),
$$\ODOT{Q} \equiv { dQ \over d\kappa} = \dot{Q} 
\>{d\tau \over d\kappa}\ . \eqno(23)$$
The relationship of the new affine parameter $d\kappa$ to the proper
time $d\tau$ is 
easily derived by dividing (21) by $d\tau$ or $d\kappa$, noting
$d\tau^2=dx^\mu dx_\mu$,
$${d\tau \over d\kappa}\equiv \left( 1 - {\dot{a}^{\mu\nu}
\dot{a}_{\mu\nu} \over 2 c^2 \lambda^2}\right)^{-1/2} =
\sqrt{1 + {\ODOTS{a}{\mu\nu}\ODOTL{a}{\mu\nu} \over
2 c^2 \lambda^2 } } \ . \eqno(24)$$

We interpret the spin to be the canonical momenta conjugate to the
bivector coordinate,
$$S_{\mu\nu} \equiv \lambda^2 \PV{\cal L\ \ }{\ODOTS{a}{\mu\nu}} =
m_0\!\ODOTL{a}{\mu\nu} = m\>\dot{a}_{\mu\nu}\ , \eqno(25)$$
$$p_{\mu} \equiv \PV{\cal L\ }{\ODOTS{x}{\mu}} =
m_0\!\ODOTL{x}{\mu} = m\>\dot{x}_{\mu}\ . \eqno(26)$$
These definitions of the momenta satisfy the Dixon equation (15).
When these momenta are reparameterized in terms of
the more familiar proper time, they have spin enhanced mass:
$m=m_0 d\tau/d\kappa$, consistent with (16).

Its easy to see that the Lagrangian (22) is invariant under
the polydimensional coordinate rotation (between vectors and
bivectors), generated by the four arbitrary
parameters $\delta \phi^\alpha$ of the automorphism transformation (8),
$$\delta x^\alpha = \lambda^{-1} \delta
\phi^\mu \> a_\mu^{\ \alpha}\ , \eqno(27)$$
$$\delta a^{\mu\nu} = \delta\phi^\mu \> x^\nu 
-\delta\phi^\nu \> x^\mu\ . \eqno(28)$$
Noether's theorem associates with this symmetry transformation a 
new set of constants of motion,
$$Q_\mu =\PV{\cal L \ }{\ODOTS{x}{\alpha}}\> \PV{x^\alpha}{\phi^\mu}
+\HALF \PV{\cal L \ \ }{\ODOTS{a}{\alpha\beta}}\>
\PV{a^{\alpha\beta}}{\phi^\mu\ } = a_\mu^{\ \alpha}\> p_\alpha 
+ S_{\mu\beta}\> x^\beta\ . \eqno(29)$$
Taking the derivative of (29) with respect to the affine parameter
yields the familiar Weysenhoff condition,
$$p_\mu\> S^{\mu\nu} = 0\ . \eqno(30)$$
This is quite significant, because usually (30) is imposed at the onset
by fiat, while we have provided an actual derivation based on the new
automorphism symmetry of the Lagrangian!

\section*{4.  General Poly-Covariance}
In general we find that particles will deviate from geodesics due to
contributions from derivatives of the basis vectors with respect
to the new bivector coordinate.  Further, in classical mechanics
the variation of the velocity is no longer equal to the velocity of
the variation.  This leads to a new derivation of the Papapetrou
equations (Papapetrou, 1951) describing the motion of spinning
particles in curved space.

\subsection*{4.1.  Covariant Derivatives in the Clifford Manifold}
The total derivative of a basis vector with respect to the new
affine parameter (24) must by the chain rule contain a derivative
with respect to the bivector coordinate,
$$\ODOT{\bf e}_\mu \equiv {d \EB{\mu} \over d \kappa} =
\ODOTS{x}{\sigma} \PD{\EB{\mu}}{x^\sigma} + \HALF 
\ODOTS{a}{\alpha\beta} \PD{\EB{\mu}\ }{a^{\alpha\beta}}\ . \eqno(31)$$
Our ans\"atze is (Pezzaglia, 1999) that the bivector derivative obeys,
$$\PD{\EB{\mu}\ }{a^{\alpha\beta}}=
\left[ \partial_\alpha,\partial_\beta \right]\EB{\mu}
-\tau^\sigma_{\alpha\beta}\>\partial_\sigma \EB{\mu}
=\left( R_{\alpha\beta\mu}^{\ \ \ \ \nu} - \tau^\sigma_{\alpha\beta}
\>\Gamma_{\sigma\mu}^\nu \right)\EB{\nu}\ , \eqno(32)$$
where $\tau^\sigma_{\alpha\beta}$ is the torsion,
$\Gamma_{\sigma\mu}^\nu $ the Cartan connection and 
$R_{\alpha\beta\mu}^{\ \ \ \ \nu}$ the Cartan curvature.

We can factor out the basis vectors by defining the covariant derivative,
$${\partial \ \ \over \partial x^\mu }
\left(p^\nu \EB{\nu} \right) = \EB{\nu} \COV{\mu}\>p^\nu \equiv\EB{\nu}
\left( \partial_\mu p^\nu + p^\sigma\>\Gamma^\nu_{\mu\sigma}
\right)\ , \eqno(33)$$
$${\partial \ \ \over \partial a^{\alpha\beta} }
\left(p^\nu \EB{\nu} \right) = \EB{\nu} 
\left[ \COV{\alpha},\COV{\beta} \right] p^\nu
\equiv \EB{\nu} \left( R_{\alpha\beta\mu}^{\ \ \ \ \nu}\>p^\mu -
\tau^\sigma_{\alpha\beta}\>\COV{\sigma}\>p^\nu \right)\ . \eqno(34)$$
From these definitions it is clear than the covariant derivatives of
the basis vectors vanish as usual.

The parallel transport of the conserved canonical momenta 
generates autoparallels in the Clifford manifold,
$$0={d \over d\kappa } \left( \EB{\mu} p^\mu\right)=
\EB{\mu}\left(\ODOTS{x}{\sigma}\COV{\sigma} +\HALF
\ODOTS{a}{\alpha\beta} \left[\COV{\alpha},\COV{\beta} \right]
 \right)p^\mu\ , \eqno(35)$$
$$0={d \over d\kappa } \left( \EB{\mu\nu}
S^{\mu\nu}\right)=
\EB{\mu\nu}\left(\ODOTS{x}{\sigma}\COV{\sigma} +\HALF
\ODOTS{a}{\alpha\beta} \left[\COV{\alpha},\COV{\beta} \right]
 \right)S^{\mu\nu}\ , \eqno(36)$$
where $\EB{\mu\nu}\equiv \EB{\mu}\wedge\EB{\nu}$.
Substituting (33) and (34), the above equations provide a new
derviation of the Papapetrou equations of motion for spinning
particles (Papapetrou, 1951).  Ours however are more general as they
include torsion and all the higher order terms.  In contravariant form,
$$0=\ODOTS{p}{\mu} +\left(\ODOTS{x}{\sigma}
\>\Gamma_{\sigma\nu}^{\ \mu} + \HALF \ODOTS{a}{\alpha\beta}
\>R_{\alpha\beta\nu}^{\,\prime\quad\mu}\right) p^\nu\ ,\eqno(37)$$
$$0=\ODOTS{S}{\rho\omega} +\delta^{\rho\omega}_{\lambda\sigma}
\left(\ODOTS{x}{\alpha}
\>\Gamma_{\alpha\nu}^{\ \lambda} + \HALF \ODOTS{a}{\alpha\beta}
\>R_{\alpha\beta\nu}^{\,\prime\quad\lambda}
\right) S^{\nu\sigma},\eqno(38)$$
$$R_{\alpha\beta\nu}^{\,\prime\quad\mu}\equiv
R_{\alpha\beta\nu}^{\ \ \ \ \mu}-\tau_{\alpha\beta}^\sigma
\>\Gamma_{\sigma\nu}^{\ \mu}\ . \eqno(39) $$

\subsection*{4.2.  An-Holonomic Mechanics}
It has been a long-standing unsolved problem to derive the Papapetrou
equations from a simple Lagrangian.
We succeed where so many others have failed because of our definition
of the new affine parameter, the form of the Lagrangian (22)
and by noting that the introduction of the
bivector coordinate has made the system an-holonomic.  Consider the
variation of the Lagrangian,
$$\delta{\cal L} = \PV{\cal L\ }{x^\alpha}\>\delta x^\alpha +
\PV{\cal L\ \ }{\ODOTS{x}{\alpha}}\>\VARO{x}{\alpha}
+\HALF \PV{\cal L\ \ }{a^{\alpha\beta}}\>\delta a^{\alpha\beta}+
\HALF \PV{\cal L\ \ }{\ODOTS{a}{\alpha\beta}}
\>\VARO{a}{\alpha\beta}\ . \eqno(40)$$
To get the equations of motion, the terms proportional to variations
of velocities must be rewritten in terms of variations of the coordinates.
Integrating the second term on the right by parts,
$$\PV{\cal L}{\ODOTS{x}{\alpha}}\VARO{x}{\alpha}=p_\alpha
\>\VARO{x}{\alpha}=
{d \over d\kappa} \left(p_\alpha\>\delta x^\alpha \right) -
\ODOTL{p}{\alpha}\>\VARO{x}{\alpha}
+p_\alpha \left(\VARO{x}{\alpha} - {d \over d\kappa}
\delta x^\alpha \right)\ . \eqno(41)$$
The leading term on the right does not contribute to the equations of
motion (the path is varied with fixed endpoints).

It is usually assumed in most undergraduate texts that the velocity of
the variation is equal to the variation of the velocity such that the
last term of (41) vanishes.  This is no longer necessarily true when
the coordinate system is anholonomic as is our case with
bivector coordinates and path dependent basis vectors.  We assert
that in general the following is valid,
$$\delta \left( \ODOTS{x}{\mu}\>\EB{\mu} \right) =
{d \over d\kappa} \left( \delta x^\mu\>\EB{\mu} \right)\ , \eqno(42)$$
$$\delta \left( \ODOTS{a}{\alpha\beta}\>\EB{\alpha}\wedge\EB{\beta}
\right) = {d \over d\kappa} \left( \delta a^{\alpha\beta} 
\>\EB{\alpha}\wedge\EB{\beta} \right)\ . \eqno(43)$$
A lengthy proof involving anholonomic coordinate transformations will
appear in Pezzaglia (1999).  In principle the derivation is an
extension of the method introduced by Kleinert (1997) for spaces
with torsion.  Performing the variations and derivatives in the above
equations and rearranging terms gives us,
$$\left(\VARO{x}{\mu} - {d \delta x^\mu \over d\kappa\ } \right) =
\delta x^\alpha \ODOTS{x}{\beta} \tau_{\alpha\beta}^\sigma
+ \HALF \left(\delta x^\alpha \ODOTS{a}{\mu\nu} -
\ODOTS{x}{\alpha}\delta a^{\mu\nu} \right)
 R_{\mu\nu\alpha}^{\,\prime\quad\sigma}\ , \eqno(44)$$
$$\left(\VARO{a}{\mu\nu} - {d \delta a^{\mu\nu}\over d\kappa\ }\right) =
\delta^{\lambda\sigma}_{\omega\nu} \left[ \Gamma^\omega_{\alpha\mu}
\left(\ODOTS{x}{\alpha}\delta a^{\mu\nu} - \ODOTS{a}{\mu\nu}
\delta x^\alpha \right) + {1 \over 4}
R_{\alpha\beta\mu}^{\prime\ \ \,\omega} \left(\ODOTS{a}{\mu\nu}
\delta a^{\alpha\beta} - \ODOTS{a}{\alpha\beta}
\delta a^{\mu\nu} \right) \right] \ . \eqno(45)$$
The first term on the right in (44) involving the torsion follows
Kleinert (1997), the rest are new.  Substituting (44) into (41) and
back into (40) and doing the same for the (45), collecting terms
proportional to $\delta x^\mu$, we obtain the anholonomic form of
the Euler-Lagrange equations of motion,
$$\PV{\cal L\ }{x^\mu} - \ODOTL{p}{\mu} + p_\lambda
\,\ODOTS{x}{\alpha}\,\tau^\lambda_{\alpha\mu} + \left(
\HALF p_\lambda\>R_{\alpha\beta\mu}^{\prime\ \ \ \lambda}
+S_{\omega\beta}\>\Gamma^\omega_{\mu\alpha} \right)
\>\ODOTS{a}{\alpha\beta} = 0\ , \eqno(46)$$
where $R^\prime$ is defined (39).  The first two terms are the
standard, the third appears in Kleinert (1997), the rest are new.
Performing the derivative on the Lagrangian (22) we recover the
covariant form of the Papapetrou equations (37).  A parallel
construction will yield the spin equation (38).

\subsection*{4.3.  Metamorphic Covariance}
Our Lagrangian (22) is invariant under {\it local} automorphism
transformations, where in general the $\phi^\mu$ of (8), (27) and (28)
can be position dependent upon a path integral of a gauge field,
$$\phi^\nu(x^\alpha) = \int^{x^\alpha} 
B^\nu_{\ \mu}(y^\sigma)\>dy^\mu \ . \eqno(47)$$
This would imply that the connection of a basis vector would become
{\it geometamorphic}\/ (Pezzaglia, 1998a),
e.g. under parallel transport a vector will turn a plane,
$$\partial_\sigma \EB{\nu} = \EB{\alpha}\, \Gamma^\alpha_{\sigma\nu}
+\EB{\mu}\wedge\EB{\nu}\,B^\mu_{\ \sigma}\ . \eqno(48)$$
Obviously this would have impact on equations (31) through (46)
of this paper.

Equation (48) is the classical analog to Crawford's (1994) spin
covariant covariant derivative for the Dirac equation derived
from generalized automorphism transformations of the Dirac algebra,
$$\COV{\mu}=\partial_\mu + i\left( eA_\mu + \gamma^5 a_\mu \right) 
+\gamma_\nu \left(\HALF B^\nu_{\ \mu} + \gamma^5\,i\,b^\nu_{\ \mu}
\right) +\HALF \gamma_{\alpha\beta}\,C^{\alpha\beta}_{\ \mu}
\ , \eqno(49)$$
$$\left(-i\hbar \gamma^\mu \COV{\mu} 
-m c \right) \psi =0\ . \eqno(50)$$
The gauge field $B^\mu_{\ \sigma}$ is the same in (48) and (49).

The Dirac equation is obtained more or less by factoring (4) into a
linear operator and replacing the momentum by the gauge-covariant
derivative: $p_\mu \rightarrow -i\hbar \COV{\mu}$.  We propose
that a generalized equation might be derived from
factoring the Dixon equation (15), and associating the commutator
derivative with the spin
operator $S_{\mu\nu} \rightarrow -i\hbar \lambda^2
\left[\COV{\mu}, \COV{\nu}\right]$.
Thus we postulate the form,
$$\left(-i\hbar \gamma^\mu \COV{\mu} -i \hbar {\lambda \over 2}
\gamma^{\alpha\beta}\left[\COV{\alpha},\COV{\beta} \right]
-m_0 c \right) \psi =0\ . \eqno(51)$$
Certainly one could include higher order triple commutator 
derivatives.  In flat space with all but the electromagnetic gauge
field $A_\mu$ suppresed in (49), the bivector (commutator)
derivative will introduce an anomalous magnetic moment interaction
which provides a possible interpretation of the constant 
$\lambda$.

\section*{5.  Summary}
In introducing {\it Dimensional Democracy}\/ we have given the bivector
a coordinate and show its utility in the treatment of the classical
spinning particle problem.  This system is invariant under
{\it polydimensional}\/ transformations which reshuffle geometry such
that `what is a vector' is {\it dimensionally relative}\/ to the
observer's frame.  A 
fundamentally new action principle has been introduced which can
accomodate anholonomic systems with torsion and spin.  Most
important, the principles proposed have potential broad
applications beyond the examples in this paper.

\section*{References}
\noindent Baylis, W. (1999). {\it Electrodynamics, A Modern 
Geometric Approach}\/, Birkh\"auser, Boston.

\medskip
\noindent Chisholm, J.S.R., and Farwell, R.S. (1991). Clifford
Approach to Metric Manifolds, in {\it Proceedings on the Winter
School on Geometry and Physics, Srni, 6-13 January, 1990}, Supplemento di
Reconditi del Circulo Matematico di Palermo, Serie {\bf 2},
no.\ 26 (1991), 123-133.

\medskip
\noindent Crawford, J.P. (1994). Local automorphism invariance:
Gauge boson mass without a Higgs particle, {\it Journal of 
Mathematical Physics}\/, {\bf 35}, 2701-2718.

\medskip
\noindent Dixon, W.G. (1970). Dynamics of extended bodies in general
relativiy, I. Momentum and angular momentum, {\it Proceedings of the
Royal Society at London}\/, {\bf A314}, 499-527.

\medskip
\noindent Kleinert, H. (1997). Nonabelian Bosonization as a
Nonholonomic Transformation from Flat to Curved Field Space,
{\it Annals of Physics}\/, {\bf 253}, 121-176; hep-th/9606065.

\medskip
\noindent Lounesto, P. (1997). {\it Clifford Algebras and Spinors}\/,
University Press, Cambridge.

\medskip
\noindent Papapetrou, A. (1951). Spinning test-particles in
general relativity, I, {\it Proceedings of the Royal Society at
London}\/, {\bf A209}, 248-258.

\medskip
\noindent Pezzaglia, W.M. (1998a). Polydimensional Relativity, A
Classical Generalization of the Automorphism Invariance Principle,
in {\it Clifford Algebras and Their Applications in Mathematical
Physics, Proceedings of fourth conference, Aachen 1996}\/, V. Dietrich,
K. Habetha and G. Jank eds., Kluwer, Dordrecht, 305-317;
gr-qc/9608052.

\medskip
\noindent Pezzaglia, W.M. (1998b). Physical Applications of a
Generalized Clifford Calculus, in {\it Dirac Operators in Analysis
(Pitman Research Notes in Mathematics, Number 394)}\/,
J. Ryan and D. Struppa eds., Longman Science \& Technology, 191-202;
gr-qc/9710027.

\medskip
\noindent Pezzaglia, W.M. (1999), Dimensionally Democratic Calculus
and Principles of Polydimensional Physics, to appear in 
{\it Proceedings of the Fifth Conference on Clifford Algebras
and their Applications in Mathematical Physics, Ixtapa, Mexico 1999,
Volume 1, Clifford Algebras and Mathematical Physics}\/, 
R. Ablamowicz and B. Fauser eds., (in preparation).

\end{document}